\documentclass[prl,aps,twocolumn,showpacs,preprintnumbers,amsmath,amssymb]{revtex4}%showkeys
\def\<{\langle}\def\>{\rangle}
\def\set#1{{\sf #1}}\def\Bnd#1{\set{B(#1)}}
\def\map#1{{\cal{#1}}}%\def\map#1{{\mathrm{#1}}}
\def\sH{\set{H}}\def\sK{\set{K}}\def\Tr{\operatorname{Tr}}
\def\Span{\set{Span}}\def\Ker{\set{Ker}}\def\Rng{\set{Rng}}
\def\){\rangle\!\rangle}\def\({\langle\!\langle}
\def\transp#1{{#1}^\tau}\def\Cmplx{\mathbb  C}
\def\partransuno#1{{#1}^{\tau_1}}\def\partransdue#1{{#1}^{\tau_2}}
\def\n#1{|\!|#1|\!|}\def\dim{\hbox{dim}}
\def\d{\operatorname{d}}
\def\eg{e. g. }
\begin{document}
\title{Imprinting a complete information about a quantum channel on
  its output state} \author{Giacomo Mauro D'Ariano}
\email{dariano@unipv.it} \altaffiliation[Also at: ]{Department of
  Electrical and Computer Engineering, Northwestern University,
  Evanston, IL 60208} \author{Paoloplacido Lo Presti}
\email{lopresti@unipv.it} \affiliation{{\em Quantum Optics and
    Information Group}, Istituto Nazionale di Fisica della Materia,
  Unit\`a di Pavia} \homepage{http://www.qubit.it}
\affiliation{Dipartimento di Fisica ``A. Volta'', via Bassi 6, I-27100
  Pavia, Italy} \date{\today} 
\begin{abstract}
  We introduce a novel property of bipartite quantum states, which we
  call {\em faithfulness}, and we say that a state is faithful when 
  acting with a channel on one of the two quantum systems, the output
  state carries a complete information about the channel.
  The concept of faithfulness can also be extended to sets of 
  states, when the output states patched together carry a
  complete imprinting of the channel.
\end{abstract}
\maketitle 
When a quantum system enters a quantum channel, its state transforms
according to a linear, trace-preserving, and completely positive map
$\map{E}$ \cite{note1}. The input-output evolution of the state $\rho$
can be written in a so-called Krauss's form \cite{Kraus83a}
\begin{equation}
  \label{krauss}
  \rho\rightarrow\map{E}(\rho)=\sum_n K_n\rho K_n^\dag\;,
\end{equation}
where $K_n$ are operators on the Hilbert space $\sH$ of the
quantum system, and satisfy the completeness relation
$\sum_n K_n^\dag K_n=I$, in order to preserve the trace
of $\rho$. 
It is natural now to pose the question: is it possible to recover the
channel $\map{E}$ from the output state $\map{E}(\rho)$? 

\begin{figure}[hbt]
  \begin{center}
    \setlength{\unitlength}{1000sp}
    \begin{picture}(8745,3219)(931,-3565)
      {\thicklines \put(5401,-1261){\oval(1756,1756)}}
      {\put(1801,-1261){\line( 1, 0){2700}}}
      {\put(6301,-1261){\vector( 1, 0){3300}}}
      {\put(1801,-3361){\vector( 1, 0){7800}}}
      \put(2026,-2611){\makebox(0,0)[b]{$R$}}
      \put(5401,-1486){\makebox(0,0)[b]{$\map{E}$}}
      \put(9676,-2536){\makebox(0,0)[b]{$R_\map{E}$}}
    \end{picture}
  \end{center}
  \caption{The input state $R$ is called {\em faithful} when the 
    correspondence between the output state $R_\map{E}\doteq
    \map{E}\otimes\map{I}(R)$ and the quantum channel $\map{E}$ is
    one-to-one, namely the output state $R_\map{E}$ carries a complete
    imprinting of the quantum channel $\map{E}$.}\label{gener}
\end{figure}
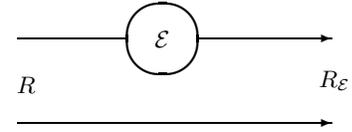

It is clear that a single input state cannot be sufficient, and one
actually needs to vary the input state over a somewhat ``complete''
set in order to recover the channel \cite{macca}, and this is the
basis of the so called {\em quantum process tomography}
\cite{nielsen}. In Ref.  \cite{tomochannel} it has been shown how a
fixed maximally entangled state can be used to recover the channel
$\map{E}$, with the entangled state playing the role of all possible
input states in a {\em quantum parallel} fashion.  The key feature of
the method is that only one of the two entangled systems is impinged
into the channel, whereas the other is left untouched, as in Fig.
\ref{gener}. The result is that for an unnormalized maximally entangled
input, such as
\begin{equation}
  |I\)=\sum_l |l\>\otimes|l\>, 
\end{equation}
where we used the notation $|A\)=\sum_{ij} A_{ij}|i\>\otimes|j\>$ for
vectors $|A\)\in\sH\otimes\sH$ for a fixed orthonormal basis
$\set{b}=\{|i\>\otimes|j\>\}$ for $\sH\otimes\sH$, the corresponding
unnormalized output
\begin{equation}
  S_\map{E}\doteq\map{E}\otimes \map{I} (|I\)\(I|)\label{S}
\end{equation}
is in one-to-one correspondence with the quantum channel $\map{E}$,
the inverse relation being
\begin{equation}
  \map{E}(\rho)=\Tr_2[I\otimes\transp{\rho}\,S_\map{E}],
  \label{oneone}
\end{equation}
where $\transp{O}$ denotes transposition of the operator $O$ with
respect to the same basis $\set{b}$. The above one-to-one
correspondence is an application of the Gelfand-Naimark-Segal
representation (GNS) \cite{Murphy} of C${}^*$-algebras: a simple proof
can be found in Ref. \cite{clon_cov}. Notice that we have kept the
input vector unnormalized, in order to have the correspondence valid also for
infinite dimensions (the correspondence actually holds between completely
bounded maps and bounded operators).

The question is now: is it possible to keep the correspondence between
output state $R_\map{E}$ and channel $\map{E}$ one-to-one also using a
non maximally entangled input state $R$?  As we will show in this
letter, the answer is affirmative. Moreover, quite surprisingly there
are even input {\em separable states} which work as well. We will call
the input states with such property {\em faithful}, namely we will use
the following definition: {\em a state $R$ is {\em faithful} when the
  output $R_\map{E}\doteq\map{E}\otimes\map{I}(R)$ from a quantum
  channel $\map{E}$ is in one-to-one correspondence with $\map{E}$}.
In other words, $R_\map{E}$ carries a complete imprinting of the
quantum channel $\map{E}$.

Using the spectral decomposition $R=\sum_l|A_l\)\(A_l|$ for the input
state $R$, we can write
\begin{eqnarray}
  R_\map{E}&=&\map{E}\otimes\map{I}(R)=
  \sum_l I\otimes\transp{A}_l S_\map{E}I\otimes A_l^*
  \nonumber\\ &\equiv&\map{E}\otimes\map{R}(|I\)\( I|),
  \label{R_E}
\end{eqnarray}
where $\map{R}(\rho)=\sum_l\transp{A}_l \rho A_l^*$, and $O^*$ denotes
the complex conjugated of the operator $O$ with respect to the basis
$\set{b}$. From Eqs. (\ref{S}) and (\ref{R_E}) we can see that the
{\em faithfulness of $R$ is equivalent to the invertibility of the map
$\map{R}$}, since whenever the map $\map{R}$ is invertible the output
state $R_\map{E}$ will be in one-to-one correspondence with $S_\map{E}$, and thus with
the channel $\map{E}$. For a pure state $R\equiv |A\)\(A|$
faithfulness means simply that the operator $A$ is invertible.  In
other words: {\em a pure bipartite state is faithful iff it has
  maximal Schmidt's number}. More generally, for mixed states
$R=\sum_l|A_l\)\(A_l|$, the full map 
$\map{R}(\rho)=\sum_l\transp{A}_l\rho A_l^*$ must be invertible, in order
to have a one-to-one correspondence between $R_\map{E}$ and
$\map{E}$.
\begin{figure}[hbt]
  \begin{center}
    \setlength{\unitlength}{1000sp}
    \begin{picture}(14008,3936)(872,-4261)
      {\thicklines \put(3533,-1297){\oval(1756,1756)}}
      {\thicklines \put(11668,-1240){\oval(1756,1756)}}
      {\thicklines \put(11599,-3346){\oval(1756,1756)}}
      \put(926,-1261){\line( 1, 0){1700}}
      \put(926,-3361){\vector( 1, 0){5800}}
      \put(4426,-1261){\vector( 1, 0){2300}}
      \put(9026,-1261){\line( 1, 0){1700}}
      \put(12526,-1261){\vector( 1, 0){2300}}
      \put(9026,-3361){\line( 1, 0){1700}}
      \put(12526,-3361){\vector( 1, 0){2300}}
      {\thicklines \put(7600,-2061){\line( 1, 0){600}}}
      {\thicklines \put(7600,-2286){\line( 1, 0){600}}}
      {\thicklines \put(7600,-2511){\line( 1, 0){600}}}
      \put(1000,-2611){\makebox(0,0)[b]{$R$}}
      \put(3526,-1486){\makebox(0,0)[b]{$\map{E}$}}
      \put(6300,-2611){\makebox(0,0)[b]{$R_\map{E}$}}
      \put(11626,-1411){\makebox(0,0)[b]{$\map{E}$}}
      \put(11551,-3511){\makebox(0,0)[b]{$\map{R}$}}
      \put(9300,-2611){\makebox(0,0)[b]{$|I\)$}}
      \put(14400,-2611){\makebox(0,0)[b]{$R_\map{E}$}}
    \end{picture}
  \end{center}
  \caption{A generally mixed state $R=\sum_l|A_l\)\(A_l|$ is 
    faithful when the map $\map{R}(\rho)=\sum_l\transp{A}_l\rho A_l^*$
    is invertible, in order to guarantee the one-to-one correspondence
    between $R_\map{E}$ and $\map{E}$.}
  \label{composit}\end{figure}
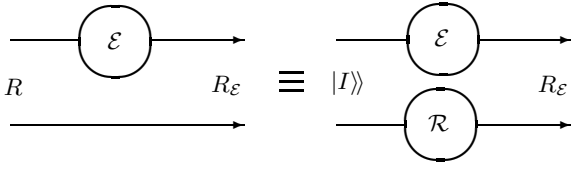
  
Let's first consider the case of finite dimensions, for simplicity. 
In terms of an orthonormal basis $\set{B}=\{B_i\}$  of operators for the
(Hilbert space of) operators $S\in\Bnd{H}$ on $\sH$ (i. e.
$\Tr[B^\dag_iB_j]=\delta_{ij}$), we have the unique expansion
$S=\sum_i\Tr[B^\dag_iS] B_i$. The map $\map{R}$ is invertible iff 
it transforms the basis into a set of linearly independent operators
$\map{R}(B_i)=\sum_j R_{ij}B_j$, where, by definition,
$R_{ij}=\Tr[B^\dag_j\map{R}(B_i)]$.  In fact the map $\map{R}$ is
invertible iff the matrix $R_{ij}$ has inverse $R_{ij}^{-1}$, and the
inverse map $\map{R}^{-1}$ is defined through its action on the basis
as $\map{R}^{-1}(B_i)=\sum_j R_{ij}^{-1}B_j$.  Then, starting from
$\map{R}(S) = \sum_{j}\Tr[B^\dag_j\map{R}(S)]B_j$, it is possible to
recover $S$ by applying $\map{R}^{-1}$ as
$S=\sum_j\Tr[B^\dag_j\map{R}(S)]\map{R}^{-1}(B_j)$.
%\begin{equation}
%S=\sum_j\Tr[B^\dag_j\map{R}(S)]\map{R}^{-1}(B_j).\label{invertR}
%\end{equation}
From the identity
\begin{equation}
  R_{ij}=\Tr[B^\dag_j\sum_l\transp{A}_lB_iA_l^*]
  =\(B_j|\sum_l\transp{A}_l\otimes A_l^\dag|B_i\),
\end{equation}
we see that an equivalent condition for the invertibility of $\map{R}$
is that the operator $\check R$ on $\sH^{\otimes 2}$
\begin{equation}
  \check R\doteq\sum_l\transp{A}_l\otimes A_l^\dag
  =\partransdue{(ER)} E\equiv \partransuno{(\partransdue{R}E)}
  \label{Rcheck}
\end{equation}
is invertible, where $E=\sum_{ij}|ij\>\<ji|$ is the swap operator, and
$O^{\tau_l}$ denotes the partial transposition of the operator $O$ on
the $l$th Hilbert space.  Actually, the correspondence
$\map{R}\leftrightarrow\check R$ preserves multiplication of maps, as
$|\map{R}(S)\)=\check R|S\)$, whence $|\map{R}^{-1}(S)\)=\check
R^{-1}|S\)$. This further clarifies the relation between the
invertibility of the map $\map{R}$ and that of the operator $\check R$
\cite{note2}.

Therefore, a bipartite state $R$ is {\em faithful} if and only if the
operator $\check R$ in Eq. (\ref{Rcheck}) is invertible.  For this
kind of states the relation $R_\map{E}=\map{E}\otimes\map{I}(R)
\leftrightarrow S_\map{E}$ is one-to-one, and the information about
the channel $\map{E}$ can be extracted from $R_\map{E}$ as follows
\begin{equation}
  \label{inversion}
  \map{E}(\rho)=\Tr_2[I\otimes\transp{\rho}\map{I}\otimes
  \map{R}^{-1}(R_\map{E})].
\end{equation}
It is clear that the set of faithful states $R$ is {\em dense} within
the set of all bipartite states. However, the knowledge of the map
$\map{E}$ from a measured $R_\map{E}$ will be affected by increasingly
large errors for $\check R$ approaching a non-invertible operator, and
{\em measures of faithfulness} should be introduced (we will analyze this problem later in this letter). Since the set of
faithful states is dense, it follows that there must be faithful
states among mixed separable states.  For example, the Werner's states
for dimension $d$
\begin{equation}
  R_f=\frac1{d(d^2-1)}[(d-f)+(df-1)E],\quad-1\leq f\leq 1,
  \label{werner}
\end{equation}
are separable for $f\geq 0$, however, they are faithful for all
$f\neq\frac1d$ , since one has $\partransdue{(E
  R_f)}=\frac1{d(d^2-1)}[(d-f)|I\)\(I| +(df-1)]$, and the singular
values of $\check R_f$ are $\frac{df-1}{d(d^2-1)}$ and $\frac{1}{d}$.
Similarly, the ``isotropic'' states
\begin{equation}
  \label{isotropic}
  R_f=\tfrac fd|I\)\(I|+
  \tfrac{1-f}{d^2-1}(I-\tfrac 1d|I\)\(I|),
\end{equation}
are faithful for $f\neq\frac1{d^2}$ and separable for $f\leq\frac1d$,
the singular values of $\check R_f$ being $\frac{d^2f-1}{d(d^2-1)}$
and $\frac fd$.

For infinite dimensions (\eg for ``continuous variables''), one needs
to restrict $\Bnd{H}$ to the Hilbert space of Hilbert-Schmidt
operators on $\sH$, and this lead to the  
problem that the inverse map $\map{R}^{-1}$ is unbounded. The
result is that we will recover the channel $\map{E}$ from the measured
$R_\map{E}$, however, with unbounded amplification of statistical
errors, depending on the chosen representation $\set{B}=\{B_j\}$, due to the 
fact that $\map{R}^{-1}(B_j)$ increases unboundedly for $j\to\infty$.
As an example, let's consider a twin beam from parametric
down-conversion of vacuum  
\begin{equation}
  |\Psi\)=\Psi\otimes I|I\),\quad
  \Psi=(1-|\xi|^2)^{\frac{1}{2}}\xi^{a^\dag a},\qquad |\xi|< 1\label{Psi}
\end{equation}
for a fixed $\xi$, $a^\dag$ and $a$, with $[a,a^\dag]=1$, denoting the
creation and annihilation operators of the harmonic oscillator
describing the field mode corresponding to the first Hilbert space in
the tensor product (in the following we will denote by $b^\dag$ and
$b$ the creation and annihilation operators of the other field mode).
The state is faithful, but the operator $\Psi^{-1}$ is unbounded,
whence the inverse map $\map{R}^{-1}$ is also unbounded. In a photon
number representation $\set{B}=\{|n\>\< m|\}$, the effect will be an
amplification of errors for increasing numbers $n,m$ of photons. 
\par Consider now the quantum channel describing the {\em Gaussian
  displacement noise} \cite{Hall}
\begin{equation}
  \map{N}_\nu (\rho)=\int_{\mathbb{C}}\frac{\d\alpha}{\pi\nu}
  \exp[-|\alpha|^2/\nu]  D(\alpha)\rho D^\dag(\alpha),
\end{equation}
where $D(\alpha)=\exp(\alpha a^\dag-\alpha^* a)$ denotes the usual
displacement operator on the phase space. The Gaussian noise is in a
sense the analogous of the depolarizing channel for infinite
dimension. The maps $\map{N}_\nu$ for varying $\nu$ satisfy the
multiplication rule $\map{N}_\nu \map{N}_\mu =\map{N}_{\nu+\mu}$,
whence the inverse map is formally given by
$\map{N}_\nu^{-1}\equiv\map{N}_{-\nu}$. Notice that, since the map
$\map{N}_\nu$ is compact, the inverse map $\map{N}_\nu^{-1}$ is
necessarily unbounded. As a faithful state consider now the mixed
state given by the twin-beam, with one beam spoiled by the Gaussian
noise, namely 
\begin{equation}
  R=\map{I}\otimes\map{N}_\nu(|\Psi\)\(\Psi|).
\end{equation}
A lengthy straightforward calculation gives the state
\begin{equation}
  R=\frac{1}{\nu}
  \Psi\otimes I\exp[-(a-b^\dag)(a^\dag -b)/\nu]\Psi^\dag\otimes I,
  \label{twinnoised}
\end{equation}
and its partial transposed
\begin{equation}
  \partransdue{R}=(\nu+1)^{-1} \Psi \otimes I 
  \left(\frac{\nu-1}{\nu+1}\right)^{\frac{1}{2}(a-b)^\dag(a-b)}
  \Psi^\dag\otimes I,
\end{equation}
where transposition is defined with respect to the basis of eigenvectors of 
$a^\dag a$ and $b^\dag b$. Since our state is Gaussian, the PPT
criterion guarantees separability \cite{Simon}, and for $\nu >1$ our
state (\ref{twinnoised}) is separable (see also
Ref. \cite{EvolTWBParis}), still it is {\em formally} faithful, since
the operator $\Psi$ and the map $N_\nu$ are both invertible. 
Notice that unboundedness of the map inversion can even wash out
completely the information on the channel in some particular chosen
representation $\set{B}=\{B_j\}$, \eg when all operators $B_j$ are out
of the boundedness domain of $\map{R}^{-1}$. This is the case, for
example, of the (overcomplete) representation
$\set{B}=\{|\alpha\>\<\beta|\}$, with $|\alpha\>$ and $|\beta\>$
coherent states, since from the identity 
\begin{equation}
  \map{N}_{\nu }(|\alpha\>\<\alpha|)=\frac{1}{\nu +1}
  D(\alpha)\left(\frac{\nu}{\nu +1}\right)^{a^\dag a}D^\dag(\alpha),
\end{equation}
one obtains
\begin{equation}
  \map{N}_{\nu}^{-1}(|\alpha\>\<\alpha|)=\frac{1}{1-\nu}
  D(\alpha)\left(1-\nu^{-1}\right)^{-a^\dag a}D^\dag(\alpha),
\end{equation}
which has convergence radius $\nu\le \frac{1}{2}$, which is the well
known bound for Gaussian noise for the quantum tomographic
reconstruction for coherent-state and Fock representations
\cite{Gausstomo}. Therefore, we say that the state is {\em formally}
faithful, however, we are constrained to representations which are
analytical for the inverse map $\map{R}^{-1}$.

In a more general framework, we can consider the faithfulness of the
bipartite state $R$ of two quantum systems described by different (in
finite dimensions) Hilbert spaces $\sH$ and $\sK$. We need now to
consider operators $A$ 
in either $\Bnd{\sK,\sH}$, $\Bnd{\sH}$, or $\Bnd{\sK}$ (in all
cases we will keep the same notation $|A\)$ for the corresponding
vector in $\sH\otimes\sK$, $\sH^{\otimes2}$, or $\sK^{\otimes2}$).
Corresponding to the state $R=\sum_l|A_l\)\(A_l|$ on $\sH\otimes\sK$
we have now the map $\map{R}(\rho)=\sum_l A_l^\tau\rho A_l^*$ from
states on $\sH$ to states on $\sK$, and the related operator $\check R= \sum_l
A_l^\tau\otimes A_l^\dag \in\Bnd{\sH^{\otimes 2},\sK^{\otimes 2}}$.
We still have $\check R|S\)=|\map{R}(S)\)$, with
$|S\)\in\sH^{\otimes 2}$, and $|\map{R}(S)\)\in\sK^{\otimes 2}$. In
order to express $\check R$ through its matrix components $R_{ij}$, we
now need to choose d an operator basis $\set{B}'=\{B'_i\}$ also for $\Bnd{\sK}$,
so that $\map{R}(B_i)=\sum_j R_{ij}B'_j$, where now
\begin{equation}
  R_{ij}=\Tr[B'^\dag_j \map{R}(B_i)]
  =\<\!\<B'_j|\sum_lA^\tau_l\otimes A_l^\dag|B_i\>\!\>\;.
\end{equation}
Then, the faithfulness of $R$, is more generally equivalent to the
left invertibility of $\map{R}$, or, equivalently, to the left
invertibility of $\check R$.

Now, if we have a set of bipartite states $\{R^{(n)}\}_1^N$ on
$\sH\otimes\sK$ which are not faithful, we can try anyway to recover the
quantum channel $\map{E}$ by patching the outputs
$R^{(n)}_\map{E}\doteq\map{E}\otimes\map{I}(R^{(n)})$ together. This
is possible if and only if the following joint state on $\sH\otimes\tilde\sK$
($\tilde\sK\doteq\sK\otimes\Cmplx^N)$ is faithful
\begin{equation}
  R_\mathrm{set}=\sum_{n=1}^N 
  p_nR^{(n)}\otimes |n\>\<n|,\label{jst}
\end{equation}
where $p_n$ are probabilities (the state $R_\mathrm{set}$ is
equivalent to a mixture the states $\{R^{(n)}\}$ in a {\em knowingly}
fashion). In this case we call the set $\{R^{(n)}\}$ a {\em
  faithful set of states}.

What can we do with an unfaithful state $R$? The map $\map{R}$ is not
invertible, and for all vectors $|S\>\!\>\in\Ker(\check R)$ we have
$\map{R}(S)=0$. However, an unfaithful state can still be
useful in recovering some quantum channels, or at least a part of
them. In fact, one can use the pseudo-inverse $\check R^\ddag$, which
allows inversion only in $\Rng(\check R^\dag)\equiv\Ker(\check
R)^\perp $, with $\check R^\ddag\check R=I-P$, $P$ being
the orthogonal projector on $\Ker(\check R)$. Correspondingly, one
defines the pseudo-inverse map $\map{R}^\ddag$ from the identity
$|\map{R}^{\ddag}(S)\)\doteq\check R^{\ddag}|S\)$, or, equivalently,
by its action on the basis for $\Bnd{\sK}$, i. e.
$\map{R}^{\ddag}(B'_i)=\sum_j R^{\ddag}_{ij}B_j$. It is clear that now
the inversion, instead of the full $S_\map{E}$, will give its projection
\begin{equation}
  \tilde{S}_\map{E}=\map{I}\otimes\map{R}^\ddag(R_\map{E})
  =\map{I}\otimes\map{Q}(S_\map{E})
  \label{sloppyinversion}
\end{equation}
where $\map{Q}=\map{R}^\ddag\map{R}=\map{Q}^2$ is the orthogonal
projection map on the operator space $\Bnd{\sH}$, also defined as
$|\map{Q}(S)\)=(I-P)|S\)=\check Q|S\)$. The partially recovered map
$\tilde{\map{E}}(\rho)=\Tr_2[I\otimes\transp{\rho}\tilde{S}_\map{E}]$
is generally not CP, and can also be written as
$\tilde{\map{E}}=\map{E}\map{Q}^*$, where $\map{Q}^*$ is the
orthogonal projection map corresponding to the operator $\check Q^*$.

The above considerations suggest a definition of a {\em number of
faithfulness} $\varphi$ as $\varphi(R)=\Tr(\check Q)$, corresponding
to the rank of $\check R$. Then, a state is faithful for
$\varphi(R)=\dim(\sH)^2$.  Notice that for 
$\varphi(R)<\dim(\sH)^2$ one can have the situation in which
$\Rng(\check R^\dag)=\Span\{|S\), S\,\text{commuting}\}$, in which
case the state $R$ allows to reconstruct completely only ``classical''
channels, with the input restricted to an abelian algebra of states.

The introduction of pseudo-inversion provides an algorithm for the
patching when using a set of states $\{R^{(n)}\}$ that 
lead to the projection maps $\{\map{Q}^{(n)}\}$.  The set is faithful
iff we can recover any operator $S\in\Bnd{\sH}$ from its projections
$\map{Q}^{(n)}(S)$, and this is possible iff, given a basis
$\{B_i\}$ for $\Bnd{\sH}$, one has
$\Span\{\map{Q}^{(n)}(B_i)\}_{i,n}=\Bnd{\sH}$.  In such circumstances,
any element of the basis can be expressed as a linear combination of
the $\map{Q}^{(n)}(B_i)$, therefore the component $\Tr[B_i^\dag S]$
of $S$ will be a linear combination of terms of the form
$\Tr[\map{Q}^{(n)}(B_i)^\dag S]$, whereas these terms can be
calculated substituting $S$ with the projection $\map{Q}^{(n)}(S)$.
Expanding $B_i=\sum_{jn}\lambda_{ij}^n\map{Q}^{(n)}(B_j)$, the
reconstruction formula for $S_\map{E}$ will be
\begin{equation}
  S_\map{E}=\sum_{ijn}\lambda_{ij}^{n*}
  \Tr_2[I\otimes\map{Q}^{(n)}(B_j)^\dag\tilde{S}^{(n)}_\map{E}]
  \otimes B_i \;.
\end{equation}

Now, let's consider the problem of how to define a measure of
faithfulness $F(R)$ of the state $R$.  As already noticed, even though
in principle any faithful state can be used to perform a tomography of
the channel $\map{E}$ \cite{tomochannel}, the experimental errors on
the measured $R_\map{E}$ are propagated to $\map{E}$ by the inversion
of the map $\map{R}$. Thus different faithful input states can produce
very different errors on the measured channel.  It is clear that all
the features producing amplification of errors are contained in the
singular values $\sigma_l$ of $\check R$: the inversion
involves multiplications by $\sigma_l^{-1}$, then the smaller are the
singular values $\sigma_l$, the larger is the error amplification, whence the
number of measurements needed to have a good reconstruction. This
suggests that a measure of faithfulness should be an increasing
function $F(R)\equiv f(\{\sigma_l\})$ of the singular values $\sigma_l
$ of $\check R$, and when the different $\sigma_l$ are equivalent
$f(\{\sigma_l\})$ must be invariant under their permutations, whence
$F(R)$ is a symmetric monotone norm of $\check R$ \cite{bhatia}. It is
clear, however, that it is unpractical to have a universal measure for
faithfulness, and its actual definition will be dictated by the
goodness criterion adopted for the reconstruction of the quantum
channel $\map{E}$. For example, the Frobenius norm $\n{\check
  R}_2=\Tr[\check R^\dag \check R]=\Tr[R^\dag 
R]$ coincides with the purity of the state $R$, and this simply shows
that, for the part of the channel that can be reconstructed, the error
is minimized for pure input state $R$.

The definitions of $F$ and $\varphi$ can be naturally extended
to sets of states $\{R^{(n)}\}$ via the introduction of the joint
state $R_\mathrm{set}$ in Eq. (\ref{jst}). It obviously follows that
any chosen degree of faithfulness $F(R)$ of a maximally entangled pure
state $R$ will be larger than the degree $F(\{R^{(n)}\})$ of any
faithful set $\{R^{(n)}\}$ of unfaithful states.

In conclusion, in this letter we have introduced a new feature of
bipartite quantum states, which we call {\em faithfulness},
corresponding to the ability of the state of carrying the complete
imprinting of a channel acting on one of the pair of quantum systems.
This property has also been extended to sets of bipartite states, when
the channel can be recovered from the corresponding output states
patched together. We have seen that there are separable states that
are faithful, and the maximally faithful states are the maximally
entangled pure states. We want to stress that the property of being
faithful is a strictly quantum feature, since a faithful state cannot
be written as the mixture of local classical (i. e. commuting) states.
This also shows how subtle is the game between the classical and
quantum natures in the correlations of a general mixed quantum state.
\par One of us (G. M. D.) acknowledges B. K\H{u}mmerer for 
pointing him the GNS and Kolmogorov's constructions, and P. Horodecki
for interesting discussions.  This work has been jointly founded by 
the EC under the program ATESIT (Contract No. IST-2000-29681), by
the US Department of Defense MURI program (Grant
No. DAAD19-00-1-0177), and by INFM project PRA-2002-CLON.


\begin{thebibliography}{10}
\bibitem{note1} A map $\map{E}$ is completely positive if it preserves
  positivity when acting locally on a bipartite state. In
  other words, upon denoting by $\map{I}$ the identical map on the
  Hilbert space $\sK$ of a second quantum system, the extended map
  $\map{ E}\otimes\map{I}$ on $\sH\otimes\sK$ must be positive for any
  extension $\sK$.
\end{thebibliography}
\end{document}